\documentclass{JHEP3} 
\usepackage{epsfig,multicol,bbm}

\def\ifm#1{\relax\ifmmode#1\else$#1$\fi}  \def\DAF{DA\char8NE}  \def\x{\ifm{\times}}
\def\pt#1,#2,{\ifm{#1\x10^{#2}}}   \def\up#1{\ifm{^{#1}}}   \def\dn#1{\ifm{_{#1}}}
\def\plm{\ifm{\,\pm}\,}   \def\ab{\ifm{\sim}}   \def\deg{\ifm{^\circ}}   \def\gam{\ifm{\gamma}}
  \def\to{\ifm{\rightarrow}}   \def\f{\ifm{\phi}}  \def\kl{\ifm{K_L}}
\def\ks{\ifm{K_S}}  \def\kpm{\ifm{K^\pm}}  \def\kp{\ifm{K^+}}  \def\km{\ifm{K^-}}
   \def\ko{\ifm{K^0}}   \def\pb{{\bf p}}   \def\epm{\ifm{e^+e^-}}
\def\kb{\ifm{\rlap{\kern.3em\raise1.9ex\hbox to.6em{\hrulefill}} K}}
\def\kob{\ifm{\kb\vphantom{K}^0}}  \def\sig{\ifm{\sigma}}  \def\dif{\hbox{d}}

\def\pio{\ifm{\pi^0\pi^0}}  \def\po{\ifm{\pi^0}}  \def\pic{\ifm{\pi^+\pi^-}}  \def\rmk{\rm\kern.5mm }

\def\figb#1;#2;{\parbox{#2cm}{\epsfig{file=#1.eps,width=#2cm}}}
\def\figbc#1;#2;{\cl{\figb #1;#2;}}   \def\minus{$-$}  
\def\ie{{\it\kern-1pt i.\kern-.5pt e.\kern-.2pt}}  \def\etal{{\it et al.}}
\let\cl=\centerline

\catcode`@=11 
\newdimen\z@ \z@=0pt 
\newskip\z@skip \z@skip=0pt plus0pt minus0pt
\def\m@th{\mathsurround=\z@}
\def\ialign{\everycr{}\tabskip\z@skip\halign} 
\def\eqalign#1{\null\,\vcenter{\openup\jot\m@th
  \ialign{\strut\hfil$\displaystyle{##}$&$\displaystyle{{}##}$\hfil
      \crcr#1\crcr}}\,}
\catcode`@=12 

\newcommand{\MeVc}{\ensuremath{{\rm MeV}}}

\newcommand{\Lcms}{\ensuremath{{\rm cm}^{-2}\,{\rm s}^{-1}}}
\newcommand{\Lpb}{\ensuremath{\rm pb^{-1}}}

\newcommand{\bra}[1]{\ensuremath{\langle\,#1\,|}}
\newcommand{\ket}[1]{\ensuremath{\,|#1\,\rangle}}

\newcommand{\BR}[1]{\ensuremath{{\rm BR}(#1)}}
\newcommand{\BRo}[1]{\ensuremath{{\rm BR}_0(#1)}}
\newcommand{\SN}[2]{\ensuremath{#1\times10^{#2}}}

\newcommand{\abs}[1]{\ensuremath{\left|#1\right|}}
\newcommand{\VSS}[3]{\ensuremath{#1\pm#2_{\rm stat}\pm#3_{\rm syst}}}

\def\ord#1;{\ifm{{\mathcal O}(#1)}}

\newcommand{\Eq}[1]{eq. \ref{#1}}
\newcommand{\Eqs}[1]{eqs. \ref{#1}}
\newcommand{\Fig}[1]{figure \ref{#1}}
\newcommand{\Ref}[1]{ref. \cite{#1}}

\newcommand{\Sec}[1]{section \ref{#1}}

\newcommand{\Tab}[1]{table \ref{#1}}

\newcommand{\Vus}{\ensuremath{|V_{us}|}}
\newcommand{\Vud}{\ensuremath{|V_{ud}|}}
\newcommand{\Vusd}{\ensuremath{|V_{us}/V_{ud}|}}
\newcommand{\fVus}{\ensuremath{|f_+(0)\,V_{us}|}}
\newcommand{\fVusVud}{\ensuremath{|V_{us}/V_{ud}\times f_K/f_\pi|}}
\newcommand{\fo}{\ensuremath{f_+(0)}}
\newcommand{\fkfp}{\ensuremath{f_K/f_\pi}}
\newcommand{\dSU}{\ensuremath{\delta^{\rm SU(2)}}}
\newcommand{\dEM}{\ensuremath{\delta^{\rm EM\vphantom{()}}}}
\newcommand{\Emiss}{\ensuremath{E_{\,\rm miss}}}
\newcommand{\pmiss}{\ensuremath{|{\bf p}_{\,\rm miss}|}}
\newcommand{\fphat}{\mbox{$\tilde{f}_+(t)$}}

\newcommand{\fzhat}{\mbox{$\tilde{f}_0(t)$}}

\newcommand{\lam}{\mbox{$\lambda_+$}}
\newcommand{\lamp}{\mbox{$\lambda'_+$}}
\newcommand{\lampp}{\mbox{$\lambda''_+$}}
\newcommand{\lamo}{\mbox{$\lambda_0$}}
\newcommand{\lamop}{\mbox{$\lambda'_0$}}
\newcommand{\lamopp}{\mbox{$\lambda''_0$}}

\title{\mathversion{bold}
\Vus\ and lepton universality from kaon decays with the KLOE detector}
\author{The KLOE Collaboration:\\
F.~Ambrosino,$^{e,f}$
A.~Antonelli,$^a$
M.~Antonelli,$^a$
F.~Archilli,$^a$
C.~Bacci,$^{j,k}$
P.~Beltrame,$^b$
G.~Bencivenni,$^a$
S.~Bertolucci,$^a$
C.~Bini,$^{h,i}$
C.~Bloise,$^a$
S.~Bocchetta,$^{j,k}$
F.~Bossi,$^a$
P.~Branchini,$^k$
R.~Caloi,$^{h,i}$
P.~Campana,$^a$
G.~Capon,$^a$
T.~Capussela,$^a$
F.~Ceradini,$^{j,k}$
F.~Cesario,$^{j,k}$
S.~Chi,$^a$
G.~Chiefari,$^{e,f}$
P.~Ciambrone,$^a$
F.~Crucianelli,$^h$
E.~De~Lucia,$^a$
A.~De~Santis,$^{h,i}$
P.~De~Simone,$^a$
G.~De~Zorzi,$^{h,i}$
A.~Denig,$^b$
A.~Di~Domenico,$^{h,i}$
C.~Di~Donato,$^f$
B.~Di~Micco,$^{j,k}$
A.~Doria,$^f$
M.~Dreucci,$^a$
G.~Felici,$^a$
A.~Ferrari,$^a$
M.~L.~Ferrer,$^a$
S.~Fiore,$^{h,i}$
C.~Forti,$^a$
P.~Franzini,$^{h,i}$
C.~Gatti,$^a$
P.~Gauzzi,$^{h,i}$
S.~Giovannella,$^a$
E.~Gorini,$^{c,d}$
E.~Graziani,$^k$
W.~Kluge,$^b$
V.~Kulikov,$^n$
F.~Lacava,$^{h,i}$
G.~Lanfranchi,$^a$
J.~Lee-Franzini,$^{a,l}$
D.~Leone,$^b$
M.~Martemianov,$^n$
M.~Martini,$^{a,e}$
P.~Massarotti,$^{e,f}$
W.~Mei,$^a$
S.~Meola,$^{e,f}$
S.~Miscetti,$^a$
M.~Moulson,$^a$
S.~M\"uller,$^a$
F.~Murtas,$^a$
M.~Napolitano,$^{e,f}$
F.~Nguyen,$^{j,k}$
M.~Palutan,$^a$
E.~Pasqualucci,$^i$
A.~Passeri,$^k$
V.~Patera,$^{a,g}$
F.~Perfetto,$^{e,f}$
M.~Primavera,$^d$
P.~Santangelo,$^a$
G.~Saracino,$^{e,f}$
B.~Sciascia,$^a$
A.~Sciubba,$^{a,g}$
A.~Sibidanov,$^a$
T.~Spadaro,$^a$
M.~Testa,$^{h,i}$
L.~Tortora,$^k$
P.~Valente,$^i$
G.~Venanzoni,$^a$
R.~Versaci,$^a$
G.~Xu,$^{a,m}$.
\\
\setcounter{page}{0}
\thispagestyle{empty}
\\
\llap{$^a$}Laboratori Nazionali di Frascati dell'INFN, Frascati, Italy\\
\llap{$^b$}Institut f\"ur Experimentelle Kernphysik, Universit\"at Karlsruhe, Germany\\
\llap{$^c$}Dipartimento di Fisica dell'Universit\`a, Lecce, Italy\\
\llap{$^d$}INFN Sezione di Lecce, Lecce, Italy\\
\llap{$^e$}Dipartimento di Scienze Fisiche dell'Universit\`a  ``Federico II'', Italy\\
\llap{$^f$}INFN Sezione di Napoli, Napoli, Italy\\
\llap{$^g$}Dipartimento di Energetica dell'Universit\`a ``La Sapienza'', Roma, Italy\\
\llap{$^h$}Dipartimento di Fisica dell'Universit\`a ``La Sapienza'', Roma, Italy\\
\llap{$^i$}INFN Sezione di Roma, Roma, Italy\\
\llap{$^j$}Dipartimento di Fisica dell'Universit\`a ``Roma Tre'', Roma, Italy\\
\llap{$^k$}INFN Sezione di Roma Tre, Roma, Italy\\
\llap{$^l$}Physics Department, State University of New York at Stony Brook, USA\\
\llap{$^m$}Institute of High Energy Physics of Academia Sinica,  Beijing, China\\
\llap{$^n$}Institute for Theoretical and Experimental Physics, Moscow, Russia\\
}

\preprint{}

\abstract{KLOE has measured most decay branching ratios of \ks, \kl\ and \kpm-mesons. It has also measured the \kl\ and the \kpm\ lifetime and determined the shape of the form factors involved in kaon semileptonic decays. We present in the following a description of the above measurements and a well organized compendium of all of our data, with particular attention to correlations. These data provide the basis for the determination of the CKM parameter \Vus\ and a test of the unitarity of the quark flavor mixing matrix. We also test lepton universality and place bounds on new physics using measurements of \Vus\ from $K_{\ell2}$ and $K_{\ell3}$ decays.}


\begin{document}
\overfullrule10pt
\section{Introduction}
While much emphasis is placed on the search for new physics, we still lack precise information on the validity of certain aspects of the Standard Model itself. In the Standard Model, the coupling of the $W$ boson to the weak charged current is written as
\begin{equation}
{g\over\sqrt2}W_\alpha^+(\rlap{\bf U}\raise2.2ex\hbox to.6em{\hrulefill}\kern.3em_L\,{\bf V}_{\rm CKM}\gamma^\alpha\,{\bf D}_L+\bar e_L\gamma^\alpha\nu_{e\,L}+\bar \mu_L\gamma^\alpha\nu_{\mu\,L}+\bar \tau_L\gamma^\alpha\nu_{\tau\,L})\ +\ \mbox{h.c.},
\end{equation}
where ${\bf U}^{\rm T}=(u,c,t)$, ${\bf D}^{\rm T}=(d,s,b)$ and $L$ is for lefthanded.
In the coupling above there is only one coupling constant for leptons and quarks. Quarks are mixed by the Cabibbo-Kobayashi-Maskawa matrix, ${\bf V}_{\rm CKM}$, which must be unitary. In low energy processes
the Fermi coupling constant $G_F$ is related to the gauge coupling $g$ by $G_F=g^2/(4\sqrt2\,M_W^2)$. In the early sixties only two elements of {\bf V}\dn{\rm CKM} were known. From nuclear $\beta$ decay it was known that \Vud\ab0.98 and from strangeness changing decays, \Vus\ab0.26, \cite{cabib}.

Precise measurements of leptonic and semileptonic kaon decay rates provide information about lepton universality. Combined with results from nuclear $\beta$ decay and pion decays, such measurements also provide information about the unitarity of the mixing matrix. Ultimately they tell us whether quarks and leptons do indeed carry the same
weak charge. The universality of electron and muon interactions can be tested by measuring the ratio $\Gamma(K\to\pi \mu \nu)/\Gamma(K\to\pi e \nu)$. The partial rates $\Gamma(K\to\pi e \nu)$ and $\Gamma(K\to\pi \mu \nu)$ provide measurements of $g^4\Vus^2$, which, combined with $g^4\Vud^2$ from nuclear $\beta$ decay and the muon decay rate,
test the unitarity condition $\Vud^2+\Vus^2+\abs{V_{ub}}^2=1$.

In 1983 it was already known that $\abs{V_{ub}}^2<4\times10^{-5}$ \cite{Klopfenstein:1983nz}
and today \abs{V_{ub}}\up2\ab\pt1.5,-5, \cite{PDG07}. We will therefore ignore \abs{V_{ub}}\up2 in the following.
The ratio $\Gamma(K\to\mu \nu)/\Gamma(\pi\to\mu \nu)$ provides an
independent measurement of $\Vus^2/\Vud^2$.

To perform these tests at a meaningful level of accuracy,
radiative effects must be properly included.
Strong-interactions introduce form factors, FF, which must be
calculated from first principles, or measured, whenever possible.
Finally, corrections for SU(2) and SU(3) breaking must also be included.
Recently, advances in lattice calculations have begun to catch up with
experimental progress.

\section{What needs to be measured}

\subsection{Semileptonic kaon decays}
\label{sec:vusl3}
The semileptonic kaon decay (\Fig{fig:amp})
still provides the best means for the measurement
\FIGURE[ht]{\figb kel3ff;4;
\caption{Amplitude for semileptonic kaon decay.\label{fig:amp}}}
\noindent of \Vus,because only the vector part of the weak current contributes
to the matrix element $\bra{\pi}J_\alpha\ket{K}$. In general,
\begin{equation}
\bra{\pi}\bar u\gamma_\alpha s\ket{K} = f_+(t)(P+p)_\alpha + f_-(t)(P-p)_\alpha,
\end{equation}
where $P$ and $p$ are the kaon and pion four-momenta, respectively,
and $t$ is the 4-momentum transfer squared $(P-p)^2$.
The form factors (FF) $f_+$ and $f_-$ appear because pions and kaons are
not point-like particles and also reflect both SU(2) and SU(3)
breaking. Lorentz invariance requires the FFs to be functions only of $t$ and therefore only of the total pion energy, $E_\pi$. Since $t=(P-p)^2=M^2+m^2-2ME_\pi$, $t$ depends only, linearly, on $E_\pi$.

Introducing the scalar FF $f_0(t)$, the matrix element above
is written as
\begin{equation}
\langle\pi(p)|\bar u\gamma_\alpha s|K(P)\rangle=\fo\x
\left((P+p)_\alpha\,\fphat+(P-p)_\alpha\left(\fzhat-\fphat\right){\Delta_{K\pi}\over t}\right),
\label{eq:kl3}
\end{equation}
with $\Delta_{K\pi}=M_{K}^2-m_{\pi}^2$. Eq. \ref{eq:kl3} defines $\tilde f_0(t)$. The FFs $f_+$ and $f_0$ must have the same value at $t=0$. We have therefore factored out a coefficient \fo, so that the functions \fphat\ and \fzhat\ are both unity at $t=0$.
For vector transitions, the Ademollo-Gatto theorem \cite{AG64:f0} ensures that SU(3) breaking is second order in $m_s-m_{u,d}$. In fact, \fo\ differs from unity by only $\sim 4\%$.
Since $P-p=k+k'$, $P-p$ dotted into the lepton term gives $m_\ell\x\bar\ell\,(1-\gam_5)\,\nu$ which can be safely neglected for $K\to\pi e\nu$ decays. Only the vector FF $\tilde f_+$ therefore contributes to $K_{e3}$ decays.

Assuming lepton universality, the muon decay rate provides the value of
the Fermi constant,
$G_F=\SN{1.16637(1)}{-5}~{\rm GeV}^{-2}$ \cite{PDG07}.
The semileptonic decay rates, fully inclusive of radiation, are given by
\begin{equation}
\Gamma(K_{\ell3(\gamma)}) =
\frac{C_K^2 G_F^2 M_K^5}{192\pi^3}\,S_{\rm EW}\,\Vus^2\,\abs{\fo}^2\,
I_{K\ell}\,\left(1 + \dSU_K + \dEM_{K\ell}\right)^2.
\label{eq:Vus}
\end{equation}
In the above expression, the index $K$ denotes $\ko\to\pi^\pm$ and $\kpm\to\po$ transitions, for which $C_K^2 =1$ and 1/2, respectively. $M_K$ is the appropriate kaon mass, $S_{\rm EW}$ is the universal short-distance electroweak correction \cite{Sir82:SEW} and $\ell=e,\:\mu$. Following a common convention, $\fo \equiv f_+^{K^0\pi^-}(0)$. The mode dependence is contained in the $\delta$ terms: the long-distance electromagnetic (EM) corrections, which depend on the meson charges and lepton masses and the SU(2)-breaking corrections, which depend on the kaon species \cite{C+02:Kl3rad}.
$I_{K\ell}$ is the integral of the dimensionless Dalitz-plot density $\rho(z,\,y)$ over the physical region for non radiative decays and includes $|\tilde f_{+,\,0}(t)|^2$.  $z,\,y$ are the dimensionless pion and lepton energies (=$2E_{\pi,\ell}/M$) and $\int \rho(z,\,y) \dif z\,\dif y$=1/4 for all masses vanishing and all FF=1. $I_{K\ell}$ does not account for virtual and real radiative effects, which are included in $\dEM_{K\ell}$.

The experimental inputs to  \Eq{eq:Vus} are the semileptonic decay rates, \ie\ branching ratios (BR) and lifetimes, and the reduced form factors \fphat\ and \fzhat, whose behavior as a function of $t$ is obtained from the decay pion spectra. At the current level of experimental precision, the choice of parametrization of the form-factor $t$ dependence becomes a relevant issue, as discussed below.

If the form factors are expanded in powers of $t$ up to $t^2$ as
\begin{equation}
\tilde{f}_{+,0}(t) = 1 + \lambda'_{+,0}~\frac{t}{m^2} +
  \frac{1}{2}\;\lambda''_{+,0}\,\left(\frac{t}{m^2}\right)^2,
\label{eq:ff2}
\end{equation}
four parameters (\lamp, \lampp, \lamop\ and \lamopp) need to be
determined from the decay pion spectrum in order to be able to compute the
phase-space integral.
However, this parametrization of the form factors is problematic,
because the values for the $\lambda$s obtained from fits to the
experimental decay spectrum
are strongly correlated, as discussed in \Ref{Fra07:Kaon}.
In particular, the correlation between
\lamop\ and \lamopp\ is $-99.96\%$; that between
\lamp\ and \lampp\ is $-97.6\%$. It is therefore impossible
to obtain meaningful results using this parametrization.

Form factors can also by described by a pole form:
\begin{equation}
\tilde f_{+,0}(t) ={M_{V,S}^2\over M_{V,S}^2-t},
  \label{eq:pole}
\end{equation}
which expands to $1 + t/M_{V,S}^2 + (t/M_{V,S}^2)^2+\ldots$. It is not obvious however what vector and scalar states
should be used.

Recent \kl\dn{e3} measurements \cite{KTeV:ff,NA48:ff,KLOE+06:Ke3FF}
show that the vector form factor $f_+(t)$ is dominated by the nearest vector
$(q\bar q)$ state with one strange and one light quark (or $K\pi$ resonance,
in an older language).
The pole-fit results are also consistent with predictions from a
dispersive approach \cite{stern,sternV,pich1}.
We will therefore make use of a parametrization for the vector form factor
based on a dispersion relation twice subtracted at $t=0$ \cite{sternV}:
\begin{equation}
\fphat = \exp\left[\frac{t}{m_\pi^2}\left( \Lambda_+ + H(t)\right)\right],
  \label{eq:sternv}
\end{equation}
where $H(t)$ is obtained from $K\pi$ scattering data.
An approximation to \Eq{eq:sternv} \cite{stern,sternV} is
\begin{equation}
\fphat = 1 + \lam{t\over m^2} + \frac{\lam^2 + p_2}{2}\left({t\over m^2}\right)^2 + \frac{\lam^3 + 3p_2\lam + p_3}{6}\left({t\over m^2}\right)^3
\label{fplus}
\end{equation}
with $\lam=\Lambda_+$. $p_2$ and $p_3$ are given in \Tab{tab:disp}, second column. The approximation is valid to \ord10\up{-3}; or better.
\TABLE{
\begin{tabular}{ccc}
\hline\hline\\[-4.5mm]
       $p_n$           & \fphat        & \fzhat        \\
\hline
$p_2 \times 10^4$ & $5.84\pm0.93$ & $4.16\pm0.50$ \\
$p_3 \times 10^4$ & $0.30\pm0.02$ & $0.27\pm0.01$ \\
\hline\hline
\end{tabular}
\caption{Constants appearing in the dispersive form of vector
and scalar form factors.}
\label{tab:disp}}

The pion spectrum in $K_{\mu3}$ decay has also been measured recently
\cite{KTeV:ff,NA48:ff2,istra2}. As discussed above, there is no
sensitivity to $\lambda''_0$. All authors have fitted their data using a
linear parametrization for the scalar form factor:
\begin{equation}
\fzhat =1+\lamop {t\over m^2}.
  \label{fzero}
\end{equation}
Because of the strong correlation between $\lambda_0'$ and $\lambda_0''$,
use of the linear rather than the quadratic parametrization gives a value
for $\lambda_0'$ which is greater that the correct value by an amount equal to
about 3.5 times the value of $\lambda_0''$. To clarify this situation,
it is necessary to obtain a form for \fzhat\ with at least $t$ and $t^2$ terms
but with only one parameter.

The Callan-Treiman relation \cite{ct} fixes the
value of scalar FF at $t=\Delta_{K\pi}$ (the so-called Callan-Treiman point)
to the ratio of the pseudoscalar decay constants $f_K/f_\pi$.
This relation is slightly modified by SU(2)-breaking corrections \cite{dct}:
\begin{equation}
\tilde{f}_0(\Delta_{K\pi})=\frac{f_K}{f_\pi}\:{1\over f_+(0)}+\Delta_{\rm CT},
  \label{eq:ct}
\end{equation}
where $\Delta_{\rm CT}$ is of \ord10^{-3};. A recent parametrization
for the scalar form factor \cite{stern} allows the constraint given by the
Callan-Treiman relation to be exploited. It is a twice-subtracted
representation of the form factor at $t=\Delta_{K\pi}$ and $t=0$:
\begin{equation}
\fzhat =\exp\left[\frac{t}{\Delta_{K\pi}} (\ln C - G(t))\right],
  \label{eq:stern}
\end{equation}
 such that $C=\tilde{f}_0(\Delta_{K\pi})$ and $\tilde{f}_0(0) = 1$.
 $G(t)$ is derived from $K\pi$ scattering data.
As suggested in \Ref{stern}, a good approximation to \Eq{eq:stern} is
\begin{equation}
\fzhat = 1 + \lamo{t\over m^2} + \frac{\lamo^2 + p_2}{2}\left({t\over m^2}\right)^2 + \frac{\lamo^3 + 3p_2\lamo + p_3}{6}\left({t\over m^2}\right)^3
\label{fzeroa}
\end{equation}
with $p_2$ and $p_3$ as given in \Tab{tab:disp}. The Taylor expansion gives $\ln C= \lamo \Delta_{K\pi}/m^2_\pi + (0.0398 \pm 0.0041)$. Eq. \ref{fzeroa} is quite similar to the result in \Ref{pich2}.

\subsection{\mathversion{bold}$K\rightarrow\mu\nu$ decays}
High-precision lattice quantum chromodynamics (QCD) results have
recently become available and are
rapidly improving \cite{Dav07:LP}. The availability of precise values for the
pion- and kaon-decay constants $f_\pi$ and $f_K$ allows use of a relation
between $\Gamma(K_{\mu2})/\Gamma(\pi_{\mu2})$ and $\Vus^2/\Vud^2$,
with the advantage that lattice-scale uncertainties and radiative corrections
largely cancel out in the ratio \cite{Mar04:fKpi}:
\begin{equation}
{\Gamma(K_{\mu2(\gamma)})\over\Gamma(\pi_{\mu2(\gamma)})}=%
{\Vus^2\over\Vud^2}\;{f_K^2\over f_\pi^2}\;%
{m_K\left(1-m^2_\mu/m^2_K\right)^2\over m_\pi\left(1-m^2_\mu/m^2_\pi\right)^2}\x(0.9930\pm0.0035),
\label{eq:fkfp}
\end{equation}
where the uncertainty in the numerical factor is dominantly from
structure-dependent radiative corrections and may be improved.
Thus, it could very well
be that the abundant decays of pions and kaons to $\mu\nu$ ultimately
give the most accurate determination of the ratio of \Vus\ to
\Vud.
This ratio can be combined with direct measurements of \Vud\ to obtain
\Vus.
What is more interesting, however, is to combine all information
from $\pi_{\mu2}$, $K_{e2}$, $K_{\mu2}$, $K_{e3}$, $K_{\mu3}$ and superallowed
$0^+ \to 0^+$ nuclear $\beta$ decays to
experimentally test electron-muon and lepton-quark universality,
in addition to the unitarity of the quark mixing matrix.

\section{KLOE at DA\char8NE}
The KLOE detector is operated at \DAF, the Frascati \f\ factory. \DAF\ is an $e^+e^-$ collider running at a center of mass energy $W = m_\f \ab 1019.45$ MeV.
\f\ mesons are produced with a cross section of \ab3 $\mu$b and decay mostly to charged kaon pairs (49\%) and neutral kaon pairs (34\%).

The neutral kaon pair from $\f\to\ko\kob$ is in a pure $J^{PC}=1^{--}$ state. Therefore the initial two-kaon state can be written, in the \f-rest frame, as
\begin{equation}
\eqalign{
\ket{K\kb,\ t=0}&=\left(\,\ket{\ko(\pb)\:\kob(-\pb)}-\ket{\kob(\pb)\:\ko(-\pb)}\,\right)/\sqrt2\cr
&\kern2cm\equiv\left(\,\ket{\ks(\pb)\:\kl(-\pb)} - \ket{\kl(\pb)\:\ks(-\pb)}\,\right)/\sqrt2,\cr}
\label{eq:kkp}
\end{equation}
where the identity holds even without assuming $CPT$ invariance. Detection of a \ks\ thus signals the presence of,  ``tags'', a \kl\ and vice versa. Thus at \DAF\ we have pure \ks\ and \kl\ beams of precisely known momenta (event by event) and flux, which can be used to measure absolute $K_S$ and $K_L$ branching ratios. In particular \DAF\ produces the only true pure \ks\ beam and the only \kl\ beam of known momentum. A \ks\ beam permits studies of suppressed \ks\ decays without overwhelming background from the \kl\ component. A \kl\ beam allows lifetime measurements. Similar arguments hold for $K^+$ and $K^-$ as well, although it is not hard to produce pure, monochromatic charged kaon beams.

In most of the following, kinematical variables will be needed in the kaon rest frame. At \DAF\ the collision center of mass, the \f-meson rest frame, is not at rest in the laboratory. Electrons and positrons collide at an angle of $\pi$-0.025 radians. The \f-mesons produced in the \epm\ collisions therefore move in the laboratory system toward the center of the accumulation rings with a momentum of about 13 MeV corresponding to $\beta_\phi$\ab0.013, $\gamma_\phi$=\ab1.00008.
$K$ mesons from \f-decay are therefore not monochromatic in the laboratory, \Fig{fig:p-th}. The \f\ momentum is measured run by run to high accuracy from Bhabha scattering.
\FIGURE[ht]{\figb p-theta;7;%
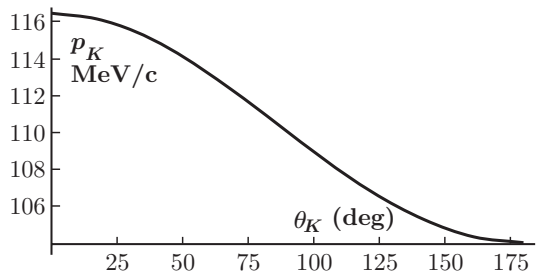
\caption{\ko-meson laboratory momentum vs angle to the $x$-axis.}\label{fig:p-th}}
The neutral kaon momentum varies between \ab104 and \ab116 MeV and is a single valued function of the angle between the kaon momentum in the laboratory and the \f\ mo\-mentum, which we take as the $x$-axis. Knowledge of the kaon direction to a few degrees allows to return to the \f-meson center of mass, Fig. \ref{fig:p-th}. The mean charged kaon momentum is 127 MeV. Boosting the laboratory measured\break quantities to whichever appropriate frame can therefore be done with great accuracy.

Because of all the above, KLOE is unique in that it is the only experiment that
can at once measure the complete set of experimental inputs, branching ratios, lifetimes and FF parameters for the
calculation of \Vus\ from both charged kaons and long lived neutral kaons. In addition KLOE is the only experiment that can measure \ks\ branching ratios at the sub-percent level.

All following discussions refer to a system of coordinates with the $x$-axis in the horizontal plane, toward the center of \DAF, the $y$ axis vertical, pointing upwards and the $z$-axis bisecting the angle of the two beam lines. The origin is at the beams interaction point, IP.

\section{The KLOE detector}
\label{sec:kloe}
At \DAF\ the mean \kl, \ks\ and \kpm\ decay path lengths are $\lambda_L = 3.4$~m, $\lambda_S = 0.59$~cm and $\lambda_\pm = 95$~cm. A detector with a radius of \ab2~m is required to define a fiducial volume for the detection of \kl\ decays with a geometrical efficiency of \ab30\%. Figure \ref{fig:kloe} shows the vertical cross section of the KLOE detector in the $y,\:z$ plane.
\FIGURE[ht]{\figbc kloe;8;\caption{Vertical cross section of the KLOE detector.\label{fig:kloe}}}%
Because the radial distribution of the \kl\ decay points is essentially uniform within this volume, tracks must be well reconstructed independently of their angles of emission. In addition, the decay points of neutral particles to photons (e.g., $\po\to\gamma\gamma$) must be localized.
To observe rare \ks\ decays and \kl\ks\ interference with no background from \ks\to\kl\ regeneration, a decay volume around the interaction point with $r>15\lambda_S$ must remain in vacuum. Material within the sensitive volume must be kept to a minimum to control regeneration, photon conversion, multiple scattering and energy loss for low-momentum charged particles. The beam pipe
surrounds the interaction point, IP,  with a sphere of 10 cm inner diameter, with walls 0.5 mm thick made of a Be-Al
sintered compound. This sphere provides a vacuum path \ab8\x\ the \ks\ {\em amplitude} decay length, effectively avoiding all \ks\to\kl\ regeneration, see fig. \ref{bpipe}.
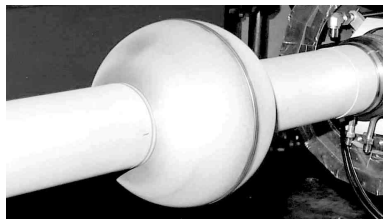
\FIGURE[htb]{\figb beampipe;5;\caption{Beam pipe at the interaction point.\label{bpipe}}}%

The detector consists principally of a large drift chamber (DC) surrounded
by a hermetic electromagnetic calorimeter (EMC). A superconducting coil
surrounding and supporting the calorimeter provides an axial magnetic field
of 0.52~T.

The DC is 3.3~m long, with inner and outer radii of 25 and 200 cm, respectively. It contains 12,582 drift cells arranged in
58 stereo layers uniformly filling the sensitive volume, for a total of 52,140 wires.
The chamber uses a gas mixture of 90\% helium and 10\% isobutane.
This reduces regeneration and multiple scattering within the
chamber, while providing good spatial resolution (150~$\mu$m).
Tracks from the origin with $\theta > 45\deg$ are
reconstructed with $\sigma(p_\perp)/p_\perp\leq$0.4\% and two-track vertices within
the sensitive volume are reconstructed with a
position resolution of \ab3~mm.
Signals from groups of 12 adjacent wires on each layer are added and digitized
providing measurements of specific ionization. This allows
identification of \kpm\ tracks by $dE/dx$ alone.
A full description of the design and operation of the chamber can
be found in \Ref{KLOE+02:DC}.

The calorimeter is built with cla\-dded, 1 mm diameter scintillating fibers embedded in 0.5-mm-thick lead foils. The foils are imprinted with grooves just large enough to accommodate the fibers and some epoxy, without compressing the fibers thus preventing damage to the fiber-cladd\-ing interface. The epoxy provides structural strength and also removes light traveling in the cladding. Many such layers are stacked, glued and pressed, resulting in a material with a radiation length $X_0$ of 1.5~cm and an electromagnetic sampling fraction of \ab13\%. This material is shaped into modules 23~cm thick (\ab$15X_0$).

24 modules of trapezoidal cross section are arranged in azimuth to form the calorimeter barrel and an additional 32 modules of square or rectangular cross section are wrapped around each of the pole pieces of the magnet yoke to form the endcaps.
The unobstructed solid-angle coverage of the calorimeter as viewed from the origin is \ab94\% of 4$\pi$. The fibers run parallel to the axis of the detector in the barrel, while they are vertical in the endcaps, and are read out at both ends with a granularity of $4.4\times4.4~{\rm cm}^2$ by a total of 4880 photomultiplier tubes, PM.

The PM signals provide the magnitude and time of energy deposits in the EMC. Deposits close in space and time are combined in clusters. Cluster energies are measured with a resolution of $\sigma_E/E = 5.7\%/\sqrt{E\ {\rm(GeV)}}$, as determined with the help of the DC using radiative Bhabha events.
The time resolution is $\sigma_t = 57~{\rm ps}/\sqrt{E\ {\rm (GeV)}}$ in quadrature with a constant term of 140~ps, as determined from radiative $\phi$ decays. The constant term results largely from the uncertainty on the collision time ($t_0$) arising from the length of the \DAF\ bunches.
The constant contribution to the relative time resolution as determined using $2\gamma$ events is \ab100~ps. Cluster positions are measured with resolutions of 1.3~cm in the coordinate transverse to the fibers and, by timing, of $1.2~{\rm cm}/\sqrt{E\ {\rm(GeV)}}$ in the longitudinal coordinate. These characteristics enable the $2\gamma$ vertex in $\kl\to\pic\po$ decays to be localized with $\sigma \approx 2$~cm along the \kl\ line of flight, as reconstructed from the tagging \ks\ decay. The calorimeter is more fully described in \Ref{KLOE+02:EmC}.

The data used for the measurements discussed in this paper were collected with a calorimeter trigger \cite{trigNIM}
requiring two energy deposits above a threshold of 50~MeV in the EMC barrel or 150~MeV in the endcaps.
The KLOE trigger also implements logic to flag cosmic-ray events, which are recognized by the presence of two energy deposits
above 30~MeV in the outermost calorimeter layers. For most KLOE data taking, such events were rejected after partial
reconstruction by an online software filter.

At a luminosity of $10^{32}$~\Lcms, events are recorded at \ab2200~Hz. Of this rate, \ab300~Hz are from \f\ decays.
Raw data, reconstructed data and Monte Carlo (MC) events are stored in a tape library. Every run is reconstructed quasi-on-line, after a complete calibration of the entire detector and measurements of the \DAF\ parameters using the immediately preceding run. For a detailed description of the data acquisition, calibration, online and offline
systems, see \cite{DAQNIM,offlineNIM}.

In 2001--2002, KLOE collected an integrated luminosity of 450~\Lpb, corresponding to approximately 140 million tagged \ks\ decays, 230 million tagged \kl\ decays and 340 million tagged \kpm\ decays.
\section{Kaon decay rate measurements}

Equation~(\ref{eq:Vus}) relates \Vus\ to the semileptonic kaon decay rates fully inclusive of radiation.
One problem that consistently plagues the interpretation of older branching ratio measurements is the lack of clarity about accounting for radiative contributions.
All of our measurements of kaon decays with charged particles in the final state are fully inclusive of radiation. Radiation is automatically accounted for in the acceptance correction. All our MC generators incorporate radiation as described in \Ref{Gat06:rad}.

\subsection{\mathversion{bold}\kl\ decays}

We search for \kl\ decays using a beam tagged by detection of
$\ks\to\pic$ decays. The \pic\ decays observed near the origin count
the number of \kl\ mesons, providing the direction and
momentum of each. We have used this technique to measure the BRs for the
four main \kl\ decay modes, as well as the \kl\ lifetime
\cite{KLOE+06:BR,KLOE+05:tauL}.

Once the $\ks\to\pic$ decay is observed,
we identify $\kl\to 3\po$ decays by the presence of multiple photons reaching the calorimeter with arrival times consistent with a unique origin along the known \kl\ flight path. We detect \kl\ decays to charged modes ($\pi e\nu$, $\pi\mu\nu$ and $\pic\po$) primarily by the observation of two tracks forming a vertex along the \kl\ path. We distinguish different decay modes by use of a single variable: the smaller absolute value of the two possible values of $\Delta_{\mu\pi} = \pmiss - \Emiss$, where \pmiss\ and \Emiss\ are the missing momentum and energy in the \kl\ decay, evaluated assuming the decay particles are a $\pi^+\mu^-$ or a $\pi^-\mu^+$ pair. Figure \ref{fig:depmiss} shows an example of a $\Delta_{\mu\pi}$ distribution.%
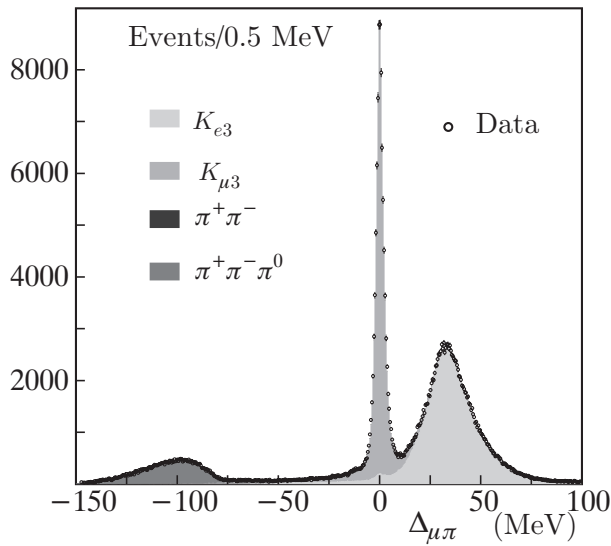
\FIGURE{\parbox{10cm}{\centerline{\figb depmiss;8;}} \caption{Distribution of $\Delta_{\mu\pi}$ for a subsample of \kl\ decays leaving two tracks in the DC.\label{fig:depmiss}}}%
We obtain the numbers of $K_{e3}$, $K_{\mu3}$ and $\pic\po$ decays by fitting the $\Delta_{\mu\pi}$ spectrum with the appropriate MC-predicted shapes. A total of approximately 13 million tagged \kl\ decays (328~\Lpb) are used for the measurement of the BRs \cite{KLOE+06:BR}.

Since the geometrical efficiency for detecting \kl\ decays in the fiducial volume chosen depends on the \kl\ lifetime $\tau_L$, so do the values of the four BRs, according to:
\begin{equation}
\BR{\kl\to f}/\BRo{\kl\to f} = 1 + 0.0128~\mbox{ns}^{-1}\:\left(\tau_L -
\tau_{L,\,0}\right),
\label{eq:fvdep}
\end{equation}
where BR\dn0 is the value of the branching ratio evaluated for a value $\tau_{L,\,0}$ of the \kl\ meson lifetime.
Our values of BR\dn0 for each mode and for a reference value of the lifetime $\tau_{L,\,0}= 51.54$ ns, the 1972 measurement of the \kl\ lifetime \cite{vosb}, are listed in \Tab{tab:klbr1}.
\TABLE{\begin{tabular}{lccccc}
\hline\hline
Parameter & Value & \multicolumn{4}{c}{Correlation coefficients} \\
\hline
\BRo{K_{e3}}          & 0.4049(21) & 1       &         &         &   \\
\BRo{K_{\mu3}}        & 0.2726(16) & $+0.09$ & 1       &         &   \\
\BRo{3\pi^0}          & 0.2018(24) & $+0.07$ & $-0.03$ & 1       &   \\
\BRo{\pi^+\pi^-\pi^0} & 0.1276(15) & $+0.49$ & $+0.27$ & $+0.07$ & 1 \\
\hline\hline
\end{tabular}
\caption{KLOE measurements of principal $K_L$ BRs assuming
$\tau_L = 51.54$~ns.}
\label{tab:klbr1}}
The four relations defined by \Eq{eq:fvdep}, together with the condition that the sum of all \kl\ BRs must equal unity, allow the determination of the \kl\ lifetime and the four BR values. This is the approach that we followed in \Ref{KLOE+06:BR}. The final KLOE results, inclusive of other measurements are given below.

An additional, independent value for $\tau_L$ is obtained from the proper decay-time distribution for $\kl\to 3\pi^0$ events, \Fig{fig:tkl}, for which the reconstruction efficiency is high and uniform over a fiducial volume of
$\sim$$0.4\,\lambda_L$ \cite{KLOE+05:tauL} ($\lambda_L$\ab3.4 m, see section \ref{sec:kloe}). About 8.5 million decays are observed within the proper-time interval $6.0-24.8$~ns; from a fit to the decay distribution we obtain
$\tau_L=\VSS{50.92}{0.17}{0.25}$ ns.%
\FIGURE[htb]{\figb taukl;6.8;\caption{Proper-time distribution for $\kl\break3\po$ decays.\label{fig:tkl}}}%

This latter measurement is included together with the results of \Tab{tab:klbr1} in a fit to determine the \kl\ BRs and lifetime. Note that the results in \Tab{tab:klbr1} are obtained without use of $\tau_L$.
We also use the KLOE measurements of $\BR{\kl\to\pic(\gam)}/\BR{{\kl}_{\mu3}}$ \cite{kltopic} and $\BR{\kl\to\gamma\gamma}/\BR{\kl\to3\po}$ \cite{klgg3po}, requiring that the seven largest \kl\ BRs add to unity.
The only non-KLOE input to the fit is the 2006 PDG ETAFIT result $\BR{\kl\to\pio}/\break\BR{\kl\to\pic} = 0.4391\pm0.0013$, based on relative amplitude measurements for $K\to\pi\pi$. We adjust this value to include direct emission in the \pic\ mode. There are thus eight experimental inputs, eight free parameters and one constraint.
The results of the fit are presented in \Tab{tab:klbr2}; the fit gives $\chi^2$/dof=0.19/1 (CL=66\%).
The BRs for the ${\kl}_{e3}$ and ${\kl}_{\mu3}$ decays are determined to within
0.4\% and 0.5\%, respectively.
\TABLE[h]{\begin{footnotesize}
\begin{tabular}{lccccccccc}
\hline\hline
Parameter & Value & \multicolumn{8}{c}{Correlation coefficients} \\
\hline
\BR{K_{e3}}          & 0.4008(15)         &   1     &         &         &         &         &         &         &   \\
\BR{K_{\mu3}}        & 0.2699(14)         & $-0.31$ &   1     &         &         &         &         &         &   \\
\BR{3\pi^0}          & 0.1996(20)         & $-0.55$ & $-0.41$ &   1     &         &         &         &         &   \\
\BR{\pi^+\pi^-\pi^0} & 0.1261(11)         & $-0.01$ & $-0.14$ & $-0.47$ &   1     &         &         &         &   \\
\BR{\pi^+\pi^-}      & \SN{1.964(21)}{-3} & $-0.15$ & $+0.50$ & $-0.21$ & $-0.07$ &   1     &         &         &   \\
\BR{\pi^0\pi^0}      & \SN{8.49(9)}{-4}   & $-0.15$ & $+0.48$ & $-0.20$ & $-0.07$ & $+0.97$ &   1     &         &   \\
\BR{\gamma\gamma}    & \SN{5.57(8)}{-4}   & $-0.37$ & $-0.28$ & $+0.68$ & $-0.32$ & $-0.14$ & $-0.13$ &   1     &   \\
$\tau_L$             & 50.84(23)~ns       & $+0.16$ & $+0.22$ & $-0.14$ & $-0.26$ & $+0.11$ & $+0.11$ & $-0.09$ & 1 \\
\hline\hline
\end{tabular}
\end{footnotesize}
\caption{Final KLOE measurements of principal \kl\ BRs and $\tau_L$.}
\label{tab:klbr2}}

\subsection{\mathversion{bold}\ks\ decays}

We have measured the ratios BR($\ks\to\pi e\nu$)/BR(\ks\to\pic) separately for each lepton charge, using \f\to\kl\ks\ decays in which the \kl\ is recognized by its interaction in the calorimeter barrel.
Semileptonic \ks\ decays are identified by time of flight (TOF) of both pion and electron. Our most recent analysis \cite{KLOE+06:KSe3} gives about 13,600 signal events, expanding upon the statistics of our original measurement \cite{KLOE+02:KSe3} by a factor of 22.
Figure~\ref{fig:kse3} shows the distributions of $\Emiss - \pmiss$ for $K_{e3}$ event candidates.
This quantity is zero for signal events which contain an unobserved neutrino. The signal peak is prominent and well separated from the background.%
\FIGURE[h]{\parbox{13cm}{\figb kse3a;6;\hfill\figb kse3b;6;}
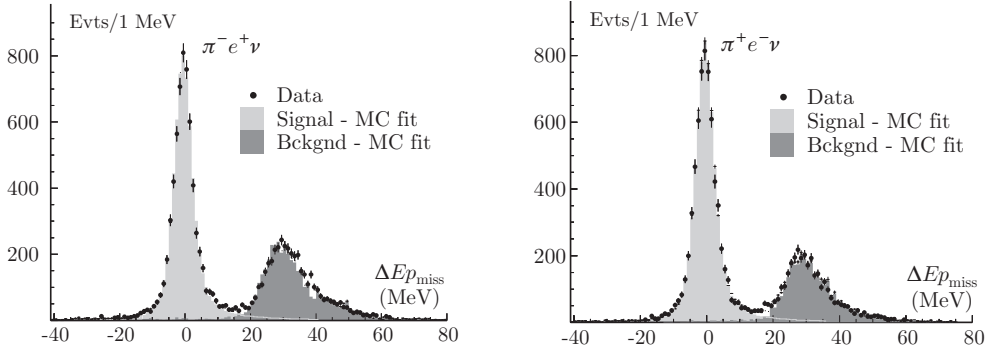
\caption{Distributions in $E_{\rm miss} - p_{\rm miss}$ (evaluated in the signal mass hypothesis) for candidate
(left) $\ks\to\pi^-e^+\nu$ and (right) $\ks\to\pi^+e^-\bar{\nu}$ events. \label{fig:kse3}}}
Combining the data for both charges we obtain BR($\ks\to\pi e\nu$)/BR(\ks\to\pic)=\SN{(\VSS{10.19}{0.11}{0.07})} {-4} \cite{KLOE+06:KSe3}. We also obtain the first measurement of the \ks\ semileptonic charge asymmetry, $A_S = \SN{(\VSS{1.5}{9.6}{2.9})}{-3}$.

In a separate analysis, we have used \ks\ decays tagged by the \kl\ interaction in the EMC barrel to measure
$\BR{\ks\to\pic}/\BR{\ks\to\pio} = 2.2549\pm0.0054$ \cite{KLOE+06:Kspp}, where this value includes a previous KLOE result \cite{KLOE+02:Kspp}. Together, these measurements completely determine the main $K_S$ BRs and give: BR($\ks\to\pi e\nu$)=\SN{(7.046\pm0.091)}{-4} \cite{KLOE+06:KSe3}.

In our evaluation of \Vus, we use the KLOE value for BR($\ks\to\pi e\nu$) together with the lifetime value $\tau_S = 0.08958\pm0.00005$~ns from the PDG fit to $CP$ parameters \cite{PDG07}. This lifetime value is mostly due to  measurements from NA48 \cite{NA48+02:tauS} and KTeV \cite{KTeV+03:CPpar}.

\subsection{\mathversion{bold}\kpm\ decays}
\label{sec:kpm}

At KLOE, $\phi\to\kp\km$ events are identified by detecting the abundant two-body decay (BR($K^\pm\to\pi^\pm\po+K^\pm\to\mu^\pm\nu)$\ab84\%) of one of the kaons. As in the analysis of neutral kaon decays, this provides tagging of the kaon of opposite charge. As noted above, the decay $K\to\mu\nu$ is of interest in its own right for the determination of \Vus.
\FIGURE[ht]{\parbox{14cm}{\centerline{\figb k2body;7;}}
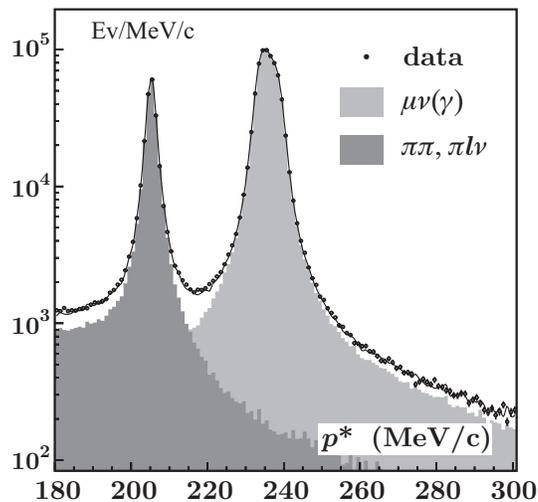
\caption{Distribution of $p^*$, the charged particle momentum in the kaon rest frame,
for $K^\pm$  decays. The solid line is the sum of all contributions, in grey, from data control samples.\label{fig:k2body}}}

Charged kaon decays are observed as ``kinks'' in a track originating at the IP. The charged kaon track must satisfy $70<p<130$ MeV. The momentum of the decay particle in the kaon rest frame, $p^*$, is  205 and 236 MeV, for $K^\pm$ decays to $\pi^\pm\po$ and $\mu^\pm\nu$, respectively.
Figure \ref{fig:k2body} shows the $p^*$ distribution obtained assuming that the decay particle is a pion. The shaded regions indicate contributions to the $p^*$ distribution as evaluated from data control samples. The peak for $K\to\mu\nu$ decays, although distorted because of the use of the wrong mass, remains clearly visible.

We measure \BR{\kp\to\mu^+\nu} (and also BR($\kp\to\pi^+\po$), \cite{erika}) using $\km\to\mu^-\bar{\nu}$ decays as tags, see \cite{KLOE+06:Km2} for details.
We obtain the numbers of $\mu\nu$ and $\pi\po$ events from the $p^*$ distribution for tagged events, as in figure \ref{fig:k2body}.
In \ab34\% of some four million tagged events, we find \ab865,000 signal events with $225 \le p^* \le 400$~\MeVc,
giving ${\rm BR}(\kp\to\mu^+\nu(\gamma)) = \VSS{0.6366}{0.0009}{0.0015}$, using efficiencies from MC and control samples.
This measurement is fully inclusive of final-state radiation (FSR), has a 0.27\% uncertainty and is independent of the charged kaon lifetime.

\FIGURE[hb]{\parbox{6cm}{\centerline{\figb kl3ch;6;}}
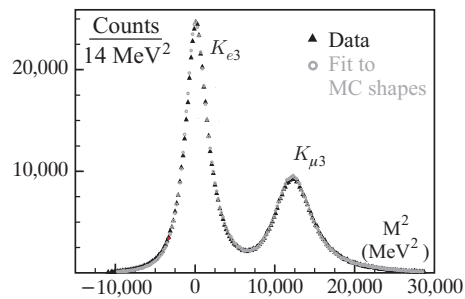
\caption{Distribution of $m^2_\ell$, from TOF information, for $K^\pm_{\ell3}$ events.\label{fig:kl3ch}}}

To measure \BR{K^\pm_{e3}} and \BR{K^\pm_{\mu3}}, we use
both $K\to\mu\nu$ and $K\to\pi\po$ decays as tags.
We measure the semileptonic BRs separately for
\kp\ and \km. Therefore, \BR{K_{e3}} and \BR{K_{\mu3}}
are each determined from four independent
measurements (\kp\ and \km\ decays; $\mu\nu$ and $\pi\po$ tags).
Two-body decays are removed from the sample by kinematics, as described above.
We then reconstruct the photons from the \po\ to reconstruct
the \kpm\ decay point. Finally, from the TOF and momentum measurement for the lepton tracks, we obtain the $m^2_\ell$ distribution shown in \Fig{fig:kl3ch}.

The number of signal events  for the two channels, plus residual background, is found fitting the $m^2_\ell$ distribution with the Monte Carlo predicted shapes.
In all, we find about 300,000 $K_{e3}$ and 160,000 $K_{\mu3}$ events.
We obtain consistent values of the BR for each decay from each of the four
subsamples; when averaged we obtain
$\BRo{K_{e3}} = (\VSS{4.965}{0.038}{0.037})\%$ and
$\BRo{K_{\mu3}} = (\VSS{3.233}{0.029}{0.026})\%$,
with a correlation of 62.7\%. Further details are given on \cite{barb}.

The above BRs are evaluated using the current world average value
for the \kpm\ lifetime, $\tau_{\pm,\,0} = 12.385$~ns \cite{PDG07}. 
The BRs depend on the value assumed for $\tau_\pm$ as
\begin{equation}
\BR{\kpm\to f}/\BRo{\kpm\to f} = 1 - 0.0364~\mbox{ns}^{-1}\:\left(\tau_\pm - \tau_{\pm,\,0}\right).
\label{eq:fvdepch}
\end{equation}
This dependence is used in our evaluation of \Vus. Both errors and correlation coefficients for the BR\dn0 values do not include contributions from the $\tau_\pm$ uncertainty.

The average value \cite{PDG07}, for the lifetime of the charged kaon is nominally quite precise: $\tau_\pm = 12.385\pm0.025$~ns. However, the consistency of the input measurements is poor: the confidence level for the average is 0.2\% and the error is scaled by 2.1. In addition the most precise result, see \cite{ott}, quotes statistical errors which are less than half of the values which follow from the given number of events. It is quite important to confirm the value of $\tau_\pm$.

At KLOE, two methods are used to reconstruct the proper decay time distribution for charged kaons.
The first is to obtain the decay time from the
kaon path length in the DC, accounting for the continuous change
in the kaon velocity due to ionization energy losses.
A fit to the proper-time distribution in the interval from 15--35~ns
($1.6\lambda_\pm$) gives the result $\tau_\pm = \VSS{12.364}{0.031}{0.031}$~ns.
Alternately, the decay time can be obtained from the precise measurement
of the arrival times of the photons from $\kp\to\pi^+\po$ decays.
In this case, a fit to the proper-time distribution in the interval from
13--42~ns
($2.3\lambda_\pm$) gives the result $\tau_\pm = \VSS{12.337}{0.030}{0.020}$~ns.
Taking into account the statistical correlation between
these two measurements ($\rho=0.307$), we obtain
the average value $\tau_\pm = 12.347\pm0.030$~ns, see \cite{massar}.
Inserting our result for $\tau_\pm$ into \Eqs{eq:fvdepch}, we obtain the
values of \kpm\ semileptonic BRs listed in \Tab{tab:kpm}, which we
use in our evaluation of \Vus.
\TABLE{\begin{tabular}{ccccc}
\hline\hline
Parameter & Value & \multicolumn{3}{c}{Correlation coefficients} \\
\hline
\BR{K_{e3}}     &  0.04972(53)   &   1     &         &    \\
\BR{K_{\mu3}}   &  0.03237(39)   & $+0.63$ &   1     &    \\
$\tau_\pm$      & 12.347(30)~ns  & $-0.10$ & $-0.09$ &  1 \\
\hline\hline
\end{tabular}
\caption{KLOE measurements of \kpm\ semileptonic decays and lifetime.}
\label{tab:kpm}}

\section{\mathversion{bold} Form factor parameters for semileptonic \kl\ decays}

To measure the $K_{e3}$ form factor parameters \cite{KLOE+06:Ke3FF},
we start from the same sample of \kl\ decays to charged particles used to measure the main \kl\
BRs. We impose additional, loose kinematic cuts and make use of time of flight
(TOF) information from the calorimeter clusters associated to the daughter
tracks to obtain better particle identification (PID).
The result is a high-purity sample of 2 million $\kl\to\pi e\nu$ decays.
Within this sample, the identification of the electron and pion tracks
is certain, so that the momentum transfer $t$ can be safely evaluated
from the momenta of the \kl\ and the daughter tracks.
We obtain the vector form factor parameters from binned log-likelihood
fits to the $t$ distribution. Using the quadratic parametrization of \Eq{eq:ff2}, we obtain
$\lamp = \SN{(\VSS{25.5}{1.5}{1.0})}{-3}$ and
$\lampp = \SN{(\VSS{1.4}{0.7}{0.4})}{-3}$,
where the total errors are correlated with $\rho = -0.95$.
Using the pole parametrization of \Eq{eq:pole}, we obtain
$M_V = \VSS{870}{6}{7}$~MeV.
Evaluation of the phase-space integral for $\kl\to\pi e\nu$ decays
gives $0.15470\pm0.00042$ using the values of \lamp\ and
\lampp\ from the first fit and $0.15486\pm0.00033$ using
the value of $M_V$ from the second; these results differ by \ab0.1\%, while
both fits give $\chi^2$ probabilities of \ab92\%.
The results we obtain using quadratic and pole fits are manifestly
consistent.

The measurement of the vector and scalar FF parameters
using $\kl\to\pi\mu\nu$ decays is more complicated.
As noted in \Sec{sec:vusl3},
there are two form factors to consider,
and since all information
about the structure of these form factors is contained in the
distribution of pion energy (or equivalently, $t$),
the correlations between FF parameters are very large.
In particular, it is not possible to measure
\lamopp\ for any conceivable number of events
\cite{Fra07:Kaon}.
In addition, at KLOE energies clean and efficient $\pi/\mu$ separation is
much more difficult
to obtain than good $\pi/e$ separation.
However, the FF parameters may also be obtained from fits to the
distribution of the neutrino energy $E_\nu$ after integration over the pion energy.
$E_\nu$ is simply the missing momentum in the
$\kl\to\pi\mu\nu$ decay evaluated in the \kl\ rest frame and
no $\pi/\mu$ identification is required to calculate it.
A price is paid in statistical sensitivity: the $E_\nu$ distribution is
related to the $t$ distribution via an integration over the pion energy.
As a result the statistical errors on the FF parameters
will be 2--3 times larger when fitting the $E_\nu$ spectrum, rather than
the $E_\pi$ spectrum. This is the case if the fit parameters
are \lamp, \lampp\ and \lamop.

For this analysis \cite{kmu3ff}, we start from the
same sample of tagged \kl\ decays to charged particles discussed
above. We impose kinematic cuts that are tighter than
those used for the $K_{e3}$ analysis and make use of shower
profile information to augment the power of the PID cuts based
on the TOF measurements for associated calorimeter clusters.
We obtain a sample of about 1.8 million $\kl\to\pi\mu\nu$ decays
with a residual contamination of \ab2.5\%, flat in $E_\nu$.
We first fit the $E_\nu$ distribution using \Eqs{eq:ff2} and
\ref{fzero} for the vector and scalar form factors.
The result of this fit is~\cite{kmu3ff}:
\begin{equation}
\eqalign{
\lamp  &= (\VSS{22.3}{9.8}{3.7}) \times 10^{-3}\cr
\noalign{\vglue-1mm}
\lampp &= (\VSS{4.8}{4.9}{1.6})  \times 10^{-3}\cr
\noalign{\vglue-1mm}
\lamop  &= (\VSS{9.1}{5.9}{2.6})  \times 10^{-3}\cr
}\kern1cm
\pmatrix{
1 & -0.97 &  0.81\cr
\noalign{\vglue-.7mm}
  &  1    & -0.91\cr
\noalign{\vglue-.7mm}
  &       &  1
}
\end{equation}
with $\chi^2/{\rm dof}=19/29$. The correlation coefficients are given as a matrix. We then combine the above results with those from our $K_{e3}$ analysis, by a $\chi^2$ fit (statistical and systematic errors are combined). We find:
\begin{equation}
\eqalign{
\lamp  &= (25.6\plm1.7) \times 10^{-3}\cr
\noalign{\vglue-1mm}
\lampp &= (1.5\plm0.8)  \times 10^{-3}\cr
\noalign{\vglue-1mm}
\lamop  &= (15.4\plm2.2) \times 10^{-3}\cr
}\kern1cm
\pmatrix{
1 & -0.95 &  0.29\cr
\noalign{\vglue-.7mm}
  &  1    & -0.38\cr
\noalign{\vglue-.7mm}
  &       &  1
}
\label{eq:ffquad}
\end{equation}
with $\chi^2/{\rm dof}=2.3/2$ and, once again,
the correlation coefficients as given in the matrix.
The same combination of $K_{e3}$ and $K_{\mu3}$ results has also
been obtained using the dispersive representations of the form factors,
\Eqs{eq:sternv} and \ref{eq:stern}, using the expansions of \Eqs{fplus} and \ref{fzeroa}.
Vector and scalar form factors are now  described by just the \lam\ and \lamo\
parameters. We find
\begin{equation}
\eqalign{
\lam  &= (\VSS{25.7}{0.4}{0.4}\pm 0.2_{\rm {param}}) \times10^{-3}\cr
\noalign{\vglue-1mm}
\lamo &= (\VSS{14.0}{1.6}{1.3}\pm 0.2_{\rm {param}}) \times10^{-3}\cr
}
\label{eq:ffdisp}
\end{equation}
with $\chi^2/{\rm dof}=2.6/3$ and a correlation coefficient of
$-0.26$. The uncertainties arising from the choice of parametrization for
the vector and scalar form factors are given explicitly.

The values of the phase-space integrals for $K_{\ell3}$ decays are listed
in \Tab{tab:ikl}, for both values
of the FF parameters, Eqs. \ref{eq:ffquad} and \ref{eq:ffdisp}, together with their fractional differences $\Delta$.
\TABLE{\begin{tabular}{ccccc}
\hline\hline
Parameters & $I(K^0_{e3})$ &$I(K^0_{\mu3})$ & $I(K^+_{e3})$ & $I(K^+_{\mu3})$     \\
\hline
\lamp, \lampp, \lamop &  0.15483(40) &  0.10271(52) & 0.15919(41) & 0.10568(54) \\
 \lam, \lamo\ &  0.15477(35) &  0.10262(47) & 0.15913(36) & 0.10559(48) \\
 \hline $\Delta$ (\%)  &  0.04        &  0.09        & 0.04        & 0.09\\
 \hline\hline
\end{tabular}
\caption{Phase-space integrals for $K_{\ell3}$ decays, using Eqs. \ref{eq:ffquad}, \ref{eq:ffdisp}. $\Delta$ is the fractional difference.}
\label{tab:ikl}}
Use of the dispersive parametrization changes the value of the phase-space integrals by at most \ab0.09\% with respect to the results obtained using quadratic and linear parametrizations
for vector and scalar form factors. The larger change observed
for $K_{\mu3}$ decays is due to the incorrect use of a linear form for the $\tilde f_0(t)$ FF which results in a larger value of \lamop\ \cite{Fra07:Kaon} and a larger integral. We shall use the dispersive results in the following. The difference between results corresponds to \ab0.03\% change in \fVus.

From the Callan-Treiman relation, \Eq{eq:ct}, we find $\fo=0.967\pm0.025$, using $\fkfp = 1.189\pm 0.007$ from a recent lattice calculation~\cite{hpqcdukqcd07:fkfp}
and $\Delta_{\rm CT} = \SN{(-3.5\pm8.0)}{-3}$, \cite{dct}.
The error on our result for \fo\ is almost entirely from the uncertainty on
our measurement of \lamo.
Our result for \fo\ can be compared with the currently most precise
lattice determination,
$\fo  = 0.9644\pm0.0049$~\cite{rbcukqcd07:f0}.

\section{{\mathversion{bold}\fVus} and lepton universality}
The SU(2)-breaking and EM corrections used to evaluate \fVus\ are summarized in \Tab{tab:radcorr}.
\TABLE{\parbox{\textwidth}{\cl{
\begin{tabular}{lcc}
\hline\hline
 Channel         & $\dSU_K$  & $\dEM_{K\ell}$ \\
\hline
$K^0_{e3}$       &          0          &   0.57(15)\%       \\
$K^0_{\mu3}$     &          0          &   0.80(15)\%       \\
$K^{\pm}_{e3}$   &         2.36(22)\%  &   0.08(15)\%       \\
$K^{\pm}_{\mu3}$ &         2.36(22)\%  &   0.05(15)\%       \\
\hline\hline
\end{tabular}}}
\caption{Summary of SU(2)-breaking and EM corrections.}
\label{tab:radcorr}}
The SU(2)-breaking correction is evaluated with ChPT to \ord p^4;, as described
in \cite{su2breaking}.
The long distance EM corrections to the full inclusive decay rate
are evaluated with ChPT to
\ord e^2p^2; \cite{su2breaking} using low-energy constants
from ref. \cite{moussallam}. The entries in the table
have been evaluated recently \cite{neufeld} and include
for the first time the values of \dEM\ for the $K_{\mu 3}$ channels for
both neutral and charged kaons.
Using all of the experimental and theoretical inputs discussed above,
the values of \fVus\ have been evaluated
for the $K_{Le3}$, $K_{L\mu3}$, $K_{Se3}$, $K^{\pm}_{e3}$,
and $K^{\pm}_{\mu3}$ decay modes,as shown in \Tab{tab:vusf} and
in \Fig{fig:f0vus}. Statistical and systematic uncertainties are added in quadrature everywhere.
\TABLE{\begin{tabular}{lcccccc}
\hline\hline
Channel & \fVus & \multicolumn{5}{c}{Correlation coefficients}  \\
\hline
 $K_{Le3}$        & 0.2155(7)  & 1    &       &      &      &    \\
 $K_{L\mu3}$      & 0.2167(9)  & 0.28 &  1    &      &      &    \\
 $K_{Se3}$        & 0.2153(14) & 0.16 &  0.08 & 1    &      &    \\
 $K^{\pm}_{e3}$   & 0.2152(13) & 0.07 &  0.01 & 0.04 &  1   &    \\
 $K^{\pm}_{\mu3}$ & 0.2132(15) & 0.01 &  0.18 & 0.01 & 0.67 & 1  \\
\hline\hline
\end{tabular}
\caption{KLOE results for \fVus.}
\label{tab:vusf}}
\FIGURE[htb]{\figb f0vus;5;
\caption{KLOE results for \fVus.\label{fig:f0vus}}}%
The five different determinations have been averaged, taking into account all correlations. We find
\begin{equation}
   \fVus = 0.2157\pm0.0006
\label{eq:f0vusresult}
\end{equation}
with $\chi^2/{\rm ndf} = 7.0/4$ (CL=$13\%$).
The values and the average are shown in \Fig{fig:f0vus}.
It is worth noting that the only external experimental input to this
analysis is the $K_S$ lifetime. All other experimental inputs are KLOE results.

To evaluate the reliability of the SU(2)-breaking correction, we compare the separate averages of \fVus\ for the neutral and
charged channels, which are $0.2159(6)$ and $0.2145(13)$, respectively.
With correlations taken into account, these values agree to within $1.1\sigma$. Alternatively, an experimental estimate of
\dSU\ can be obtained from the difference between the results for neutral and
charged kaon decays, with no SU(2)-breaking corrections applied
in the latter case.
We obtain $\dSU_{\rm exp} = 1.67(62)\%$, which is in agreement with
the value estimated from theory (\Tab{tab:radcorr}).

Comparison of the values of \fVus\ for $K_{e3}$ and $K_{\mu3 }$ modes
provides a test of lepton universality. Specifically,
\begin{equation}
r_{\mu e} \equiv \frac{\fVus^2_{\mu3, \ {\rm exp}}}
                      {\fVus^2_{e3, \ {\rm exp}}} =
                 \frac{\Gamma_{\mu 3}}{\Gamma_{e3}}\:
                 {I_{e3}\left(1+\delta_{Ke}\right)^2\over
                  I_{\mu3}\left(1+\delta_{K\mu}\right)^2},
\label{eq:leptuniv}
\end{equation}
where $\delta_{K\ell}$ stands for
$\dSU_K + \dEM_{K\ell}$.
By comparison with \Eq{eq:Vus}, $r_{\mu e}$ is equal to the ratio
$g^2_{\mu}/g^2_{e}$, with
$g_{\ell}$ the coupling strength at the $W \rightarrow \ell \nu$ vertex. In the standard model, $r_{\mu e}=1$.
Averaging between charged and neutral modes, we find
\begin{equation}
r_{\mu e} =  1.000\pm0.008.
\label{eq:unilep}
\end{equation}
The sensitivity of this result may be compared with that obtained
for $\pi \rightarrow \ell \nu$ decays,
$(r_{\mu e})_{\pi} = 1.0042\pm0.0033$~\cite{piuniv},
and for leptonic $\tau$ decays,
$(r_{\mu e})_{\tau} = 1.000\pm0.004$~\cite{DHZ06}.

\section{Test of CKM unitarity}
In the previous section, a determination of  \fVus\ from $K_{\ell3}$ decays has been obtained, with fractional accuracy of 0.28\%.
Lattice evaluations of \fo\ are rapidly improving in precision. The RBC and UKQCD Collaborations have recently obtained
$\fo = 0.9644\pm0.0049$ from a lattice calculation with $2+1$ flavors of dynamical domain-wall fermions \cite{rbcukqcd07:f0}.
Using their value for \fo, our $K_{\ell3}$ results give \Vus = 0.2237\plm0.0013.
A recent evaluation of \Vud\ from $0^+ \rightarrow 0^+$ nuclear beta decays \cite{vud07ht},
gives \Vud=0.97418\plm0.00026 which, combined with our result above, gives \Vud\up2+\Vus\up2\minus 1=\minus0.0009\plm0.0008, a result compatible with unitarity,  which is verified to \ab0.1\%.
\Fig{fig:vusvud} shows a compendium of all the KLOE results.
\FIGURE[h]{\parbox{7cm}{\figb allfit4;7;}
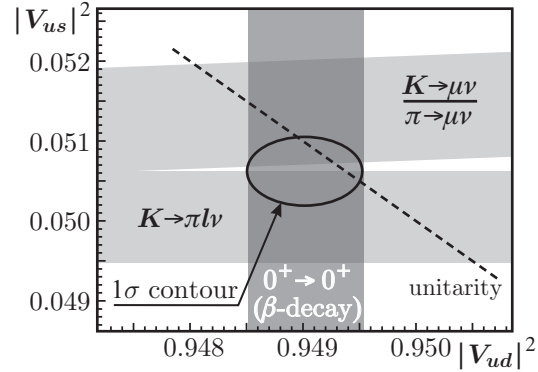
\caption{KLOE results for $|V_{us}|^2$, $|V_{us}/V_{ud}|^2$ and $|V_{ud}|^2$ from $\beta$-decay measurements, shown as 2\sig\ wide grey bands. The ellipse is the 1 \sig\ contour from the fit. The unitarity constraint is illustrated by the dashed line.}\label{fig:vusvud}}%

Additional information is provided by\break the determination of the ratio $|V_{us}/V_{ud}|$, following the approach of
\Eq{eq:fkfp}. From our measurements of BR($K_{\mu 2}$) and
$\tau_{\pm}$ and using $\Gamma(\pi_{\mu2})$ from \Ref{PDG07}, we find $\fVusVud^2 = 0.7650(33)$.
Using the recent lattice determination of \fkfp\ from the HP\-QCD/UKQCD
collaboration, \fkfp =1.189\break\plm0.007 \cite{hpqcdukqcd07:fkfp}, we
obtain $|V_{us}/V_{ud}|^2$=0.0541\break\plm0.0007. The best estimate of  $\Vus^2$ and $\Vud^2$ can be obtained from a fit to the above ratio and our result $\Vus^2$=0.05002\plm0.00057 together with the result \Vud\up2\kern1mm=\kern1mm0.9490\plm\break0.0005 from superallowed $\beta$-decays.
The fit gives $\Vus^2$ = 0.0506\plm0.0004 and $\Vud^2$ = 0.9490\kern1mm\plm\kern1mm0.0005 with a correlation of 3\%.\break The fit CL is 13\% ($\chi^2$/ndf = 2.34/1). The values obtained confirm the unitarity of the CKM quark mixing matrix as applied to the first row. We find
 $$1-\Vus^2-\Vud^2=0.0004\pm0.0007\quad(\ab0.6\sigma)$$
 \ie\ the unitarity condition is verified to \ord0.1\%;, see \Fig{fig:vusvud}. In a more conventional form, the results of the fit are:
\begin{equation}
\eqalign{|V_{us}|&=0.2249\pm0.0010\cr
|V_{ud}|&=0.97417\pm0.00026\cr}
\end{equation}
Imposing unitarity as a constraint, $\Vus^2+\Vud^2=1$, on the values above and performing a constrained fit we find
\begin{equation}
\eqalign{|V_{us}|&=0.2253\pm0.0007\cr
|V_{ud}|&=\sqrt{1-|V_{us}|^2}=0.97429\pm0.00017.\cr}
\end{equation}
The correlation is of course \minus100\% and $\chi^2/{\rm dof}$=0.46/1 corresponding to a CL of 50\%.

One should also keep in mind that while lattice results for $f_+(0)$ and $f_K/f_\pi$ appear to be converging and are quoted with small errors there is still a rather large spread between different calculations. If we were to use instead $f_+(0)=0.961\pm0.008$ as computed in \cite{L&R} and still preferred by many authors, we find $|V_{us}|=0.2258\pm0.0012$ which is less precise but satisfies more closely unitarity.

\section{Bounds on new physics from $K_{\mu 2}$ decay}
A particularly interesting test is the comparison between the
values for \Vus\
obtained from helicity-suppressed $K_{\ell2}$
decays and helicity-allowed $K_{\ell3}$ decays.
To reduce theoretical uncertainties and make use of the
results discussed above, we exploit the ratio
\BR{K_{\mu2}}/\BR{\pi_{\mu2}} and study the quantity
\begin{equation}
R_{\ell23} = \left|
  \frac{V_{us}(K_{\mu2})}{V_{us}(K_{\ell3})}\times
  \frac{V_{ud}(0^+\to0^+)}{V_{ud}(\pi_{\mu2})}
  \right|.
\end{equation}
This ratio is unity in the SM, but would be affected
by the presence of non-vanishing scalar or right-handed currents.
A scalar current due
to a charged Higgs exchange is expected to lower the value of
$R_{\ell23}$, which becomes (see \cite{IP06}):
\begin{equation}
R_{\ell23} = \left|
  1 - \frac{m^2_{K^+}}{m^2_{H^+}}\:
         \left(1 - \frac{m^2_{\pi^+}}{m^2_{K^+}}\right)\:
     \frac{\tan^2 \beta}{1 + \epsilon_0\,\tan \beta}
\right|.
\label{eq:rl23}
\end{equation}
with $\tan \beta$ the ratio of the two Higgs vacuum expectation values
in the MSSM and $\epsilon_0 \approx 0.01$ \cite{IR01}.
Any effects of scalar currents on $0^+\to0^+$ nuclear transitions
and $K_{\ell3}$ decays are expected to be insignificant and \Vus\
and \Vud\ as estimated from these modes are assumed to satisfy the
unitarity condition. A comparison of eq. \ref{eq:rl23} with experiment leads to the exclusion of some values of $m_{H^+}$, $\tan\beta$, see \Fig{fig:higgs}.
\FIGURE[t!]{\parbox{6cm}{\figb higgs;6;}
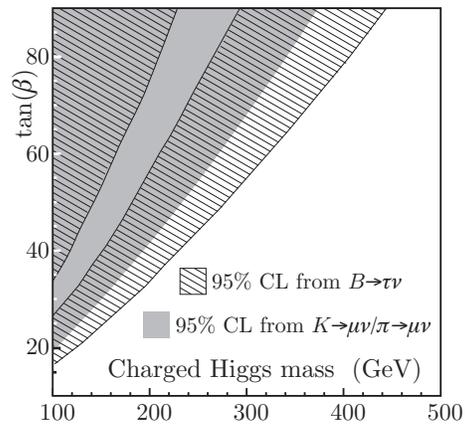
\caption{Region in the $m_{H^+}$-$\tan\beta$ plane excluded by our result
for $R_{\ell23}$; the region excluded by measurements of
$\BR{B \rightarrow \tau \nu}$ is also shown.\label{fig:higgs}}}%

To evaluate $R_{\ell23}$, we fit our experimental data on
$K_{\mu2}$ and $K_{\ell3}$ decays, using the lattice
determinations of \fo\ and \fkfp\ and the value of \Vud\
discussed above as inputs. 
We obtain
\begin{equation}
  R_{\ell23} = 1.008 \pm 0.008,
\end{equation}
which is $1\sigma$ above the standard model prediction.
This measurement places bounds on the charged Higgs mass
and $\tan \beta$.
\Fig{fig:higgs} shows the region in the
$\{m_{H^+},\ \tan\beta\}$ plane excluded at 95\% CL by our result for $R_{\ell23}$.
Measurements of $\BR{B \rightarrow \tau \nu}$~\cite{btaunu}
also set bounds on $m_{H^+}$ and $\tan\beta$, as shown in the figure.
While the $B\rightarrow \tau \nu$ data exclude an extensive region of
the plane, there is an uncovered region
corresponding to the change of sign of the correction.
This region is fully covered by our result.

\section{Conclusions}

We have measured with very good accuracy all of the main $K_S$, $K_L$ and $K^+$ BRs, the $K_L$ and $K^+$ lifetimes,
and the form factor parameters for semileptonic $K_L$ decays.
We obtain $\fVus = 0.2157\pm0.0006$ from a weighted average of the
determinations for the
$K_{Le3}$, $K_{L\mu3}$, $K_{Se3}$, $K^{\pm}_{e3}$ and $K^{\pm}_{\mu3}$ modes.
We have also tested lepton universality in $K_{\ell3}$ decays.
We obtain $r_{\mu e} = 1.000\pm0.008$, a measurement of the ratio
$g^2_{\mu}/g^2_{e}$ of the muon and electron gauge couplings.
From our measurements of the $K_{\mu2}$ decay rate,
we obtain $\fVusVud^2 =  0.7650\pm0.0033$. Our determinations of
both \fVus\ and \fVusVud\ have fractional uncertainties of
\ab0.3\% and are comparable in precision to the
present world averages \cite{PDG07}.

Using recent lattice determinations for the meson form factors,
we obtain $\Vus = 0.2237\pm0.0013$ and $\Vusd = 0.2326\pm0.0015$.
We perform a fit to combine these values with the most recent
evaluation of \Vud\ from nuclear $\beta$ decays.
The result of this fit satisfies the first-row CKM unitarity
condition to within $0.6\sigma$ and gives $\Vus = 0.2249\pm0.0010$ and
$\Vud = 0.97417\pm0.00026$, unchanged from the input value. Imposing unitarity results
in $\Vus = 0.2253\pm0.0007$ and $\Vud=\sqrt{1-\Vus^2}=0.97429\pm0.00017$ with
a correlation of \minus100\%.

Comparing the values for $\Vus$ obtained from $K_{\mu2}$ and $K_{\ell3}$
decays, we are able to exclude a large region in the $\{m_{H^+},\ \tan \beta\}$
plane. The bounds from our measurements are complementary to those
from results on $B\to\tau\nu$ decays.

\acknowledgments

We wish to acknowledge many useful discussions with Gino Isidori and Federico Mescia and thank them for help. We thank the DAFNE team for their efforts in maintaining low background running conditions and their collaboration during all data-taking. We want to thank our technical staff: G.F. Fortugno and F. Sborzacchi for their dedicated work to ensure an efficient operation of the KLOE computing facilities; M. Anelli for his continuous support to the gas system and the safety of the detector; A. Balla, M. Gatta, G. Corradi and G. Papalino for the maintenance of the electronics; M. Santoni, G. Paoluzzi and R. Rosellini for the general support to the detector; C. Piscitelli for his help during major maintenance periods. This work was supported in part by EURODAPHNE, contract FMRX-CT98-0169; by the German Federal Ministry of Education and Research (BMBF) contract 06-KA-957; by the German Research Foundation (DFG), 'Emmy Noether Programme', contracts DE839/1-4; by INTAS, contracts 96-624, 99-37; and by the EU Integrated Infrastructure Initiative HadronPhysics Project under contract number RII3-CT-2004-506078.

\newcommand{\posk}[1]{\href{http://pos.sissa.it//archive/conferences/046/#1/KAON_#1.pdf}{PoS(KAON)#1}}
\catcode`@=11
\catcode`\%=12
\catcode`\|=14
\renewcommand\jphg[3]   {\@spires{JPHGB
        {{\it J. Phys.\ }{\bf G #1} (#2) #3}}
\catcode`\%=14
\catcode`\|=12
\catcode`@=12

\bibliographystyle{JHEP}

\end{document}